\documentclass[11pt,a4paper,abstract=on]{scrartcl}

\usepackage[utf8]{inputenc}
\usepackage[english]{babel}

\usepackage[paper=a4paper,left=25mm,right=25mm,top=25mm,bottom=30mm]{geometry}
\usepackage{amsmath,amssymb,amsthm,mathrsfs,bbm,amsfonts,bm}
\usepackage{color,graphicx,cases,caption,enumerate}
\usepackage{float,multicol,authblk}
\usepackage[dvipsnames]{xcolor}
\usepackage{colortbl}
\usepackage{booktabs}
\usepackage{multirow}
\usepackage{setspace}
\usepackage{booktabs}
\usepackage{bbm}
\usepackage{capt-of}
\usepackage{array}
\usepackage[title]{appendix}
\newcolumntype{C}[1]{>{\centering\let\newline\\\arraybackslash\hspace{0pt}}m{#1}}
\usepackage{natbib}
\usepackage{placeins} 
\usepackage[breaklinks, colorlinks=true, citecolor=black, urlcolor=black, linkcolor=black]{hyperref}
\usepackage{subcaption}
\usepackage{dirtree}
\usepackage{rotating}

\title{Operational convection-permitting COSMO/ICON ensemble predictions at observation sites (CIENS)}

\author[1,2,3]{Sebastian Lerch} 
\author[1]{Benedikt Schulz}
\author[4]{Reinhold Hess}
\author[5]{Annette Möller}
\author[4]{Cristina Primo}
\author[4]{Sebastian Trepte}
\author[4]{Susanne Theis}

\affil[1]{Karlsruhe Institute of Technology, Karlsruhe, Germany}
\affil[2]{Philipps-Universität Marburg, Marburg, Germany}
\affil[3]{Heidelberg Institute for Theoretical Studies, Heidelberg, Germany}
\affil[4]{Deutscher Wetterdienst, Offenbach, Germany}
\affil[5]{Bielefeld University, Bielefeld, Germany}

\date{\today}

\begin{document}

\maketitle

\begin{abstract}
\noindent
We present the CIENS dataset, which contains ensemble weather forecasts
from the operational convection-permit\-ting numerical weather prediction model of the German Weather Service. 
It comprises forecasts for 55 meteorological variables mapped to the locations of synoptic stations, as well as additional spatially aggregated forecasts from surrounding grid points, available for a subset of these variables.
Forecasts are available at hourly lead times from 0 to 21 hours for two daily model runs initialized at 00 and 12 UTC, covering the period from December 2010 to June 2023.
Additionally, the dataset provides station observations for six key variables at 170 locations across Germany: pressure, temperature, hourly precipitation accumulation, wind speed, wind direction, and wind gusts.
Since the forecast are mapped to the observed locations, the data is delivered in a convenient format for analysis.
The CIENS dataset complements the growing collection of benchmark datasets for weather and climate modeling. A key distinguishing feature is its long temporal extent, which encompasses multiple updates to the underlying numerical weather prediction model and thus supports investigations into how forecasting methods can account for such changes.
In addition to detailing the design and contents of the CIENS dataset, we outline potential applications in ensemble post-processing, forecast verification, and related research areas.
A use case focused on ensemble post-processing illustrates the benefits of incorporating the rich set of available model predictors into machine learning-based forecasting models.
\end{abstract}

\section{Introduction}

Weather forecasts today are typically based on numerical weather prediction (NWP) models, which use systems of partial differential equations to simulate atmospheric processes. By running NWP models with varying initial conditions and/or model physics, ensemble simulations enable probabilistic forecasting. Despite continual advances \citep{Bauer2015}, NWP ensemble predictions often exhibit systematic errors, making post-processing essential for achieving accurate and reliable forecasts. Here, post-processing refers to methods that leverage past NWP forecast-observation pairs to optimally adjust future forecasts. Over the past decades, a wide variety of post-processing methods has been developed, and these now form a critical component of the forecasting workflow in national and international meteorological services \citep{VANNITSEM2018}.

A recent focus of post-processing research has been the application of modern machine learning (ML) methods; see \citet{Haupt2021, VannitsemEtAl2021} for overviews. For example, post-processing approaches based on random forests \citep{TaillardatEtAl2016}, gradient boosting \citep{Messner&2017}, and neural networks \citep{RaspAndLerch2018} have demonstrated promising results across various applications. The rapid development and growing variety of new methods clearly underscore the need for systematic comparisons and rigorous assessments of the advantages and disadvantages of these approaches. Several studies have undertaken such efforts for univariate \citep{RaspAndLerch2018, SchulzLerch2022, DemaeyerEtAl2023} and multivariate \citep{Wilks2015, PerroneEtAl2020, LerchEtAl2020, Lakatos2023} post-processing, using both simulated and real-world data. Nonetheless, there remains a critical need for comprehensive and easily accessible real-world benchmark datasets to enable fair quantitative comparisons and facilitate interdisciplinary research efforts by reducing the time-intensive process of data collection and curation \citep{DuebenEtAl2022}.

In recent years, numerous benchmark datasets for weather and climate modeling have been released, including datasets for sub-seasonal and seasonal weather forecasting \citep{Hwang2019, LenkoskiEtAl2022, Vitart2022, mouatadid2023subseasonalclimateusa} and for data-driven weather and climate prediction \citep{RaspEtAl2020, WatsonEtAl2022, WB2}. Among these, the WeatherBench 2 dataset \citep{WB2} provides global gridded NWP forecasts and corresponding reanalysis fields, making it a valuable resource for post-processing research as well \citep[see, e.g.,][]{BuelteEtAl2024}. Additionally, several recent benchmark datasets explicitly identify post-processing as a key application \citep{Haupt2021, ENS10, KimEtAl2022}.
Most closely related to our work is the recent EUPPBench dataset \citep{DemaeyerEtAl2023}, which was published as part of the activities within the post-processing working group of the European Meteorological Network (EUMETNET). EUPPBench includes two years of forecasts and 20 years of corresponding reforecasts from the European Centre for Medium-Range Weather Forecasts (ECMWF), along with station observations across Europe.

In this work, we introduce the CIENS dataset, which provides location-specific ensemble forecasts from the operational convection-permitting NWP model of the German Weather Service (Deutscher Wetterdienst, DWD), along with corresponding observations from 170 stations across Germany. 
The dataset includes ensemble forecasts of the 20 members for 55 meteorological variables mapped to the locations of the stations, as well as additional spatially aggregated forecasts from surrounding grid points available for some variables. Forecasts are available at hourly lead times from 0 to 21 hours, for two daily model runs initialized at 00 UTC and 12 UTC. 
Spanning December 2010 to June 2023, the dataset also includes station observations of six key variables: pressure, temperature, hourly precipitation accumulation, wind speed, wind direction, and wind gusts. 
Over this time period, the operational NWP model has undergone significant updates, including changes of resolution, ensemble generation mechanisms and model physics.

To the best of our knowledge, the CIENS dataset is the largest available archive of pre-processed, analysis-ready weather forecast and observation data specifically designed for station-based post-processing in terms of the temporal extent of the provided data. 
It complements existing datasets such as EUPPBench and enables addressing a wide range of research questions concerning the development and evaluation of post-processing methods. 
For instance, adapting post-processing methods to accommodate ongoing changes in the NWP model remains a challenge in operational settings \citep{Hess_2020,lang2020remember, VannitsemEtAl2021, PrimoEtAl2024}. 
The extensive time span of operational ensemble forecasts available in the CIENS dataset makes it ideally suited to address this challenge, among many others. 
Parts of the CIENS dataset have been used in previous research, primarily focused on wind gust forecasting. For example, \citet{Hess_2020,SchulzLerch2022} and \citet{PrimoEtAl2024} compare various statistical and ML-based post-processing methods, \citet{Pantillon_2018} and \citet{Eisenstein2022} investigate meteorological aspects of wind gust forecasts during severe storms, and \citet{ArnoldEtAl2024} leverage forecast and observation data for methodological advancements in forecast evaluation.

The remainder of this article is structured as follows. Section \ref{sec:data} provides a detailed description of the dataset structure, forecasts, and observations included in the CIENS dataset. Section \ref{sec:use_cases} explores potential applications of the dataset in post-processing research and other areas, while Section \ref{sec:examples} presents an exemplary use case, where ML-based post-processing methods use different sets of input variables for probabilistic wind gust forecasting are compared. The article concludes with a discussion in Section \ref{sec:conclusions}, and instructions for accessing the data and accompanying code are provided in Section \ref{sec:code_data}.

\section{CIENS dataset}\label{sec:data}

This section provides an overview of the structure and contents of the CIENS dataset. Details on the availability of code and data will be discussed in Section \ref{sec:code_data}.

\begin{sidewaystable}
	\centering
	\caption{Overview of the different parts of the CIENS dataset. The included NWP forecasts are separated between so-called standard and spatial variables. We refer to standard variables as the meteorological variables taken from the closest grid point, while the spatial variables refer to summary statistics of surrounding grid cells.  \label{tab:datasets}}
	\begin{tabular}{p{2.8cm}lp{7.7cm}c}
		\toprule
		Dataset & DOI & Content & Size \\
		\midrule
		CIENS & \href{https://dx.doi.org/10.35097/EOvvQEsgILoXpYTK}{10.35097/EOvvQEsgILoXpYTK} & ``Parent'' (or primary) dataset which serves as the official reference and links to the four parts listed below. & - \\
		\midrule 
		CIENS -- Run 00 UTC & \href{https://dx.doi.org/10.35097/zzfEJPxDILXwSNPH}{10.35097/zzfEJPxDILXwSNPH} & NWP forecasts (standard variables) of the model runs initialized at 00 UTC and observational data. & 75.9 GB \\
		CIENS -- Run 00 UTC - Spatial Variables & \href{https://dx.doi.org/10.35097/wVDXkDCGnBgFuuGt}{10.35097/wVDXkDCGnBgFuuGt} & NWP forecasts (spatial variables) of the model runs initialized at 00 UTC. & 109.7 GB \\
		CIENS -- Run 12 UTC & \href{https://dx.doi.org/10.35097/JKALdQqqLIjGUOBC}{10.35097/JKALdQqqLIjGUOBC} & NWP forecasts (standard variables) of the model runs initialized at 12 UTC. & 75.3 GB \\
		CIENS -- Run 12 UTC - Spatial Variables & \href{https://dx.doi.org/10.35097/rJZCZYljpSReTWNL}{10.35097/rJZCZYljpSReTWNL} & NWP forecasts (spatial variables) of the model runs initialized at 12 UTC. & 109.5 GB \\
		\bottomrule 
	\end{tabular}
\end{sidewaystable}

\subsection{Data structure} 

The CIENS data is provided in four parts, see Table \ref{tab:datasets} for an overview. This simplifies downloading and handling of the large dataset (with a total size of approximately 370 GB) and was necessitated by technical restrictions of the data repository, where the data is hosted (KITOpen, a central repository service at the Karlsruhe Institute of Technology). To support typical uses in the context of ensemble post-processing (see Sections \ref{sec:use_cases} and \ref{sec:examples}), we split the data by initialization times of the model runs (at 00 UTC and 12 UTC), and according to the type of meteorological variables from the NWP forecasts. Specifically, we distinguish between ensemble forecasts taken from the grid point closest to a station location, and spatial forecasts which summarize forecasts from the surrounding $11\times 11$ and $21\times 21$ grid points via their mean value and standard deviation (individually for each ensemble member). 
Note that while observations are available for all hours of the day, the (complete) observational data are only included in the CIENS -- Run 00 UTC dataset to avoid unnecessary duplicates.

The four parts of the CIENS dataset are organized in a similar manner, exemplified by the directory structure of the CIENS -- Run 00 UTC data shown in Figure \ref{fig:directory}. The forecast data is provided as daily netCDF files, with corresponding NWP ensemble predictions for all available observation stations and lead times. Corresponding observation data is provided in yearly netCDF files and can be matched to the forecast data using the provided code, see Section \ref{sec:code_data}. 

\begin{figure}    
			\small  
		\dirtree{%
			.1 CIENS -- Run 00 UTC.
			.2 observations/\DTcomment{This directory holds observation data as netCDF files organized by year.}.
			.3 obs-2010.nc.
			.3 obs-2011.nc.
			.3 ....
			.3 obs-2023.nc.
			.2 run0/\DTcomment{Directory with NWP forecast data organized in sub-folders named by year and month.}.
			.3 201012/\DTcomment{Directory with forecasts for a specific month, as netCDF files organized by day.}.
			.4 grib\_2010120800.nc\DTcomment{Forecasts are available from 8 December 2010.}.
			.4 grib\_2010120900.nc.
			.4 ....
			.4 grib\_2010123100.nc.
			.3 201101/.
			.4 grib\_2011010100.nc.
			.4 ....
			.4 grib\_2011013100.nc.
			.3 ....
			.3 202306/.
			.4 grib\_2023060100.nc.
			.4 ....
			.4 grib\_2023063000.nc\DTcomment{Forecasts are available until 30 June 2023.}.
		}
\caption{Directory structure of the CIENS -- Run 00 UTC data. \label{fig:directory}   }
\end{figure}

\subsection{NWP model forecasts}\label{sec:NWP} 

\begin{figure}
	\centering
	\includegraphics[width=\textwidth]{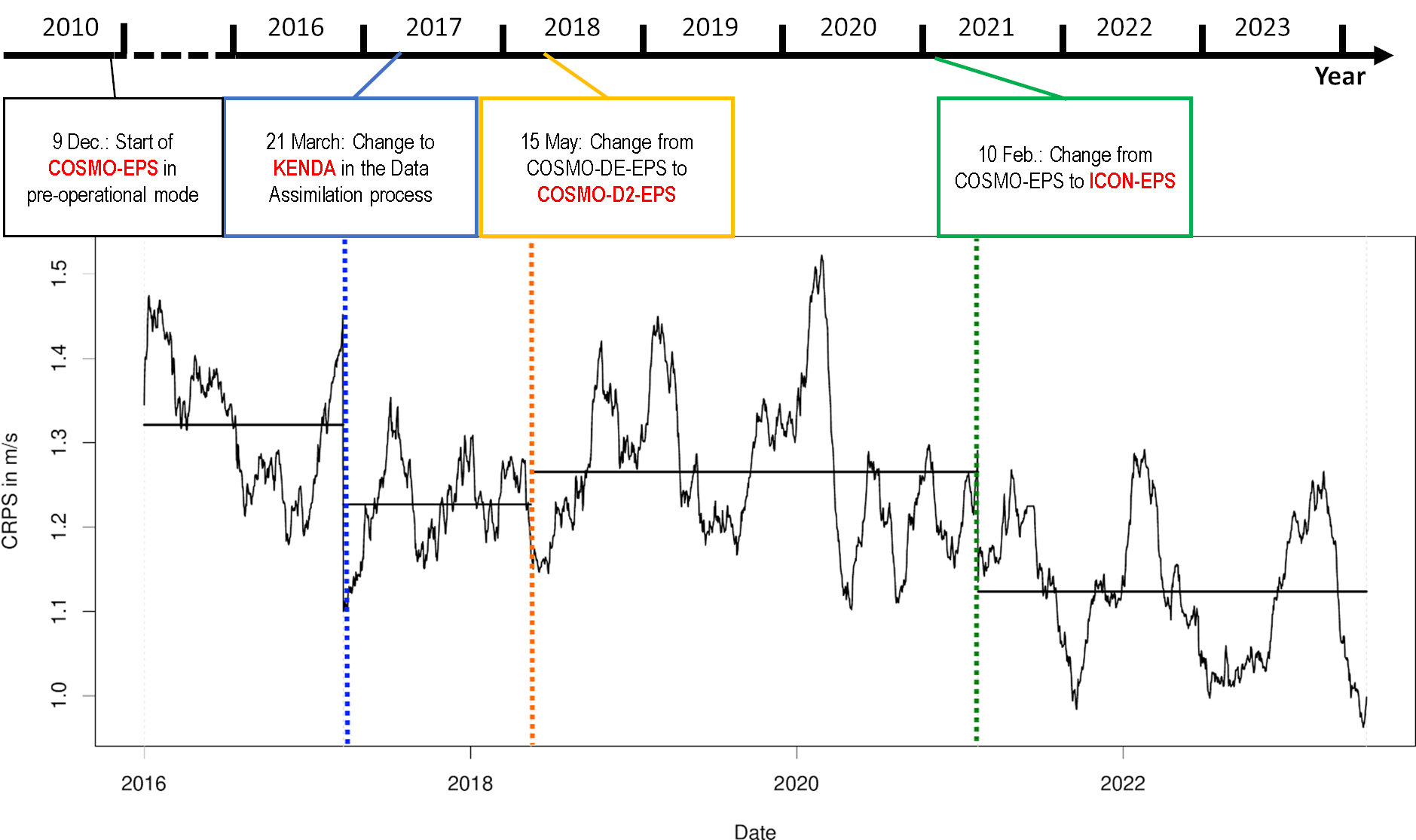}
	\caption{Overview of the most substantial changes in the NWP model underlying the CIENS forecasts indicated by the colored boxes and lines, along with the temporal evolution of the mean CRPS values (see Section \ref{sec:verification}) of the raw ensemble forecasts of wind gust with a lead time of 18 hours. The CRPS values are averaged over all station locations and smoothed with a 30-day running mean, restricted to the corresponding NWP model version. Horizontal lines indicate the mean CRPS over the corresponding period. The mean CRPS values shown here are restricted to the time period from 2016 until the end of the dataset in June 2023.}
	\label{fig_NWP_changes}  
	\bigskip
	\includegraphics[scale=.6]{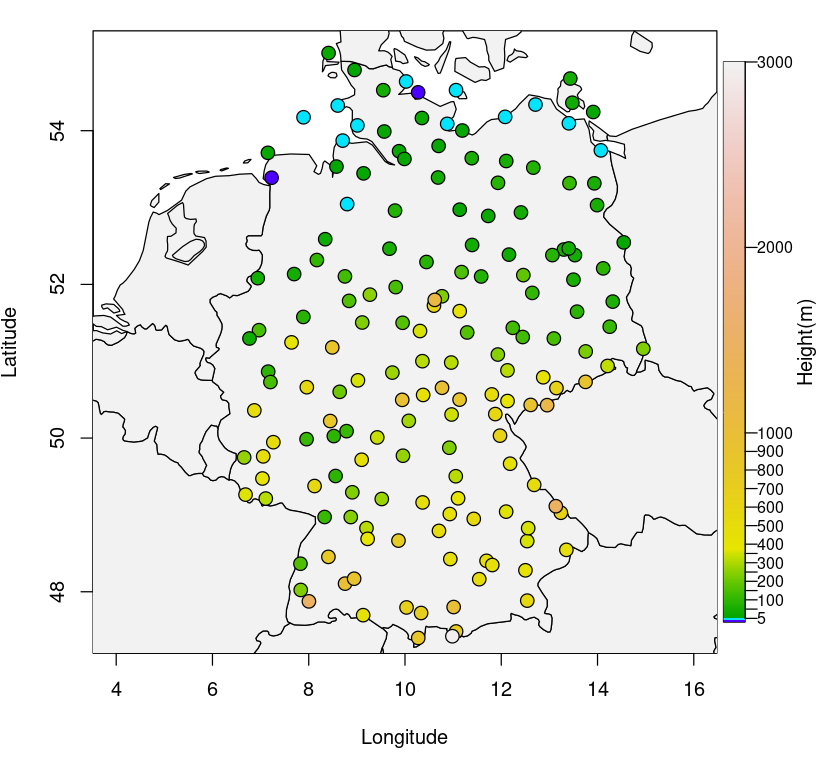}
	\caption{Map of WMO synoptic stations included in CIENS. Colors represent the station altitude (m).}
	\label{fig_stations}    
\end{figure}

The CIENS data set includes model forecasts provided by the ensemble prediction system (EPS) of the DWD from 8 December 2010 to 30 June 2023. During that period, different improvements have been made to the numerical weather prediction system, partly even resulting in different model names: COSMO-DE-EPS, COSMO-D2-EPS and ICON-D2-EPS. Figure \ref{fig_NWP_changes} shows an overview of the most relevant changes in the NWP model during the time range considered. The mean verification scores for wind gust forecasts from the NWP model shown alongside the model updates indicate that the model updates can have substantial impacts on the quality of the resulting ensemble forecasts.  

Forecast data are provided for the 00 UTC and 12 UTC runs of the ensemble forecast systems which were operational at that time, i.e., of COSMO-DE-EPS \citep{Baldauf2011,Gebhardt2011}, COSMO-D2-EPS \citep{Baldauf2018}, and ICON-D2-EPS \citep{Reinert_2021}. 
In order to generate long time series, data of COSMO-DE-EPS are used from 8 December 2010 00 UTC until 15 May 2018 12 UTC, when the first data of COSMO-D2-EPS became available. 
Among other updates, this model change included an increase in the horizontal resolution from 2.8 km to 2.2 km and an updated orography. Beginning with 10 February 2021, run 12 UTC, forecast data from the current operational ensemble system ICON-D2-EPS are used. The spatial resolution was kept constant for this model change. 
There have been numerous additional model updates, which are documented in \citet{DWD_I, DWD_II, DWD_III}. For example, at the time of writing, four model updates have occurred in the year 2024, and the latest update from 9 July 2024 comprises a revision of the wind gust parameterization and modifications to the radar data assimilation processes.
Whether such model updates will have substantial impacts on the forecast quality depends on the specific target variable of interest, along with many other factors such as the location or lead time under consideration. 

Forecasts in the CIENS dataset are available for hourly lead times from 0 to 21 hours, for each of the 20 members of the ensemble models mentioned above. The forecast model data are interpolated to 170 synoptic observation stations within Germany (see Section \ref{sec:obs}). 
The interpolation is applied separately for each ensemble member and uses data from the nearest model grid point. Furthermore, medium- and large-scale predictors are derived from the model forecasts including the spatial mean and standard deviation of 11 $\times$ 11 and 21 $\times$ 21 model grid points, respectively, around the locations of the synoptic stations, computed separately for each ensemble member.

Altogether 55 model variables are available, including near surface parameters such as 2m-temperature and dew point, wind and wind gusts in 10m height, total precipitation, cloud coverage, radiation, and many more, but also temperature, relative humidity, wind, vertical velocity and geopotential on 5 pressure levels from 500 hPa up to 1000 hPa are provided. The complete list of forecast variables is included in the CIENS repository \citep{github}.

The zonal and meridional wind components U and V of the COSMO models are rotated according to the used rotated grid. The ICON-D2-EPS uses the same rotated grid as COSMO-D2-EPS, but U and V are directed truly geographical. Therefore, in order to generate a consistent set of data, the wind components U and V of ICON-D2-EPS are rotated according to the rotated grids of the COSMO models.

\subsection{Observations at station locations}\label{sec:obs} 
%

The CIENS observation data set consists of netCDF files that include 170 European synoptic observation sites distributed within the German domain for the time range from 8 December 2010 to 30 June 2023 (see Figure \ref{fig_stations}). These station data are part of the synoptic observations distributed via the World Meteorological Observation (WMO) Global Telecommunication System (GTS), available from 2001 onward. The maximum temporal resolution is 1 or 3 hours for the standard elements.

The CIENS observation files differ from the original WMO files. An original WMO data file is written in a fixed machine-readable ASCII, however, the CIENS observations are written into netCDF files. WMO files contain all stations and all observation dates for one day with 79 elements. Not all quantities are measured at all stations and may thus be marked with missing values. The standard elements are 2m temperature, dew point, precipitation amount, mean wind speed, wind gusts, wind direction, present weather and cloud cover. In addition, there are cloud heights, sunshine duration, global radiation and many more. However, to enable a large number of observation stations covering all available variables and to minimize missing data and temporal gaps, the CIENS data only includes wind, temperature and precipitation, see Table \ref{tbl:synop_obs}. 

The only metadata included is the station identifier (WMO or national identifier), but not the station name nor any geographical information. These can be found in the World Meteorological Organization’s official repository named OSCAR/Surface\footnote{\url{https://oscar.wmo.int/surface/\#/}}.

\section{Potential uses}\label{sec:use_cases}

The comprehensive collection of forecasts and observations in the CIENS dataset enables researchers to address a wide range of questions. This section aims to present a non-exhaustive list of relevant topics in post-processing research and related fields where the CIENS dataset could serve as a valuable resource.

One primary application of the CIENS dataset in post-processing research is benchmarking both existing and novel methods for the weather variables listed in Table \ref{tbl:synop_obs} in various settings. In particular, the recent advancements and successes in ML-based post-processing underscore the need for large archives of training data. Given that a key advantage of ML models
(and other approaches, such as vine copula-based models, see, e.g., \citealp{Jobst&2023}) lies in their ability to effectively leverage information from a wide range of available predictors, the CIENS dataset—with its 55 meteorological variables from the COSMO/ICON model—provides a promising testing ground.

Specifically, it will be interesting to see whether statistical or ML-based post-processing methods can make efficient use of the additionally available spatially aggregated predictions, or whether incorporating information from all ensemble members can provide improvements over methods based on summary statistics alone \citep{hohlein_postprocessing_2024}.
Further aspects of model development in post-processing include determining optimal ways to utilize information across multiple lead times \citep{Mlakar2024} and to effectively combine multiple NWP model runs from different initialization times during model training \citep{PrimoEtAl2024}.
As noted in the introduction, frequent updates to NWP models need to be accounted for by post-processing systems, and thus pose a challenge in operational weather prediction at meteorological services \citep{VannitsemEtAl2021}. Producing a large archive of reforecasts for past dates with an updated model version would be the ideal solution for training post-processing models, but is usually infeasible in terms of the required computational resources in practice, see \citet{hamill2018practical} for a detailed discussion.
The CIENS dataset, with its extensive archive of operational forecast data encompassing several major updates, allows for detailed investigations of the effects of NWP model changes and the adaptation of post-processing methods, see Section \ref{sec:NWP}.
Another key research focus in post-processing literature has been on extreme events \citep[e.g.,][]{lerch2013comparison, williams2014comparison, Pantillon_2018, friederichs2018postprocessing}. The large volume of forecast and observation data available for variables such as wind gusts and hourly precipitation accumulation will facilitate comparative studies and targeted model development \citep{wessel2024improving} using the CIENS dataset.

\begin{table}
	\begin{center} 
		\scalebox{0.8}{
			\begin{tabular}{>{}l@{\hspace{2em}}l@{\hspace{2em}}l} 
				\toprule
				Variable & Name & Unit  \\ 
				\midrule
				wind\_speed\_of\_gust 	 & Wind gusts				& $m/s$ \\
				wind\_speed		         & Wind speed				& $m/s$ \\
				wind\_from\_direction	 & Wind direction			& $Degree$ \\
				precipitation\_amount	 & Precipitation amount (hourly)		& $kg/m^2$ \\
				air\_temperature		 & Air temperature			& $K$ \\
				air\_pressure		   & Air pressure				& $Pa$ \\
				\bottomrule 
			\end{tabular}
		}
	\end{center}
	\caption{Observed variables from the European synoptic stations included in CIENS.} 
	\label{tbl:synop_obs}	
\end{table}

In addition to univariate post-processing of ensemble forecasts for single target variables at specific locations and lead times, many applications require accurate modeling of dependencies across space, time, and variables \citep{schefzik2013uncertainty}. Consequently, recent research has increasingly focused on developing multivariate post-processing methods, including new generative ML-based models \citep{chen2024generative}, or vine copula-based methods \citep{Jobst&2024b, Jobst&2025}. The amount of target variables, locations, and lead times in the CIENS dataset provides an opportunity to expand existing benchmarking efforts, particularly through incorporating additional input predictors into multivariate post-processing models.

Beyond ensemble post-processing, the CIENS dataset also supports various other research avenues. 
For instance, it could serve as a platform for developing new verification methods for probabilistic forecasts. 
Although substantial progress has been made in both methodology and software tools \citep[for overviews, see, e.g.,][]{gneiting2007strictly, gneiting2014probabilistic, JordanEtAl2019, gneiting_etal_2023_model, allen2024weighted}, there remains a need for new approaches that address specific challenges such as extremes \citep{LerchEtAl2017, allen2023evaluating} and multivariate evaluation \citep[see, e.g.,][for a discussion from a multivariate post-processing perspective]{chen2024generative}.
Additionally, the extensive archive of data allows for a feature-based assessment of forecast quality in both raw and post-processed ensemble predictions, see, e.g., \citet{Eisenstein2022} for a study on wind gusts during winter storms.
Moreover, the CIENS forecast and observation data could be integrated with other data sources for downstream applications such as hydrological modeling or energy forecasting \citep[potentially in conjunction with post-processing, see][]{PhippsEtAl2022}.

In addition to research, the CIENS dataset could serve as valuable resource for teaching university-level courses in atmospheric sciences, statistics, or computer science, and could also be used to run forecasting competitions \citep{bracher2024learning}. Finally, the availability of a ready-to-use benchmark dataset alongside open-source software greatly simplifies data collection for student thesis projects.

\section{Usage examples for ensemble post-processing applications}\label{sec:examples}

To illustrate the usage of the CIENS data, this section presents an exemplary application for post-processing of ensemble forecasts of wind gusts. The presented analysis follows \cite{SchulzLerch2022}, but is less extensive since our main intention is to demonstrate how the CIENS data can be utilized for post-processing, and not to provide a comprehensive comparison of state-of-the-art methods. 
The example considers only the forecasts initialized at 00 UTC, for a lead time of 12 hours, at all 170 station locations. 
To take advantage of the large set of predictors available in the CIENS data, we compare a basic model that only utilizes summary statistics of the forecasts of the variable to be post-processed as inputs with an extended ML model that can process any number of input variables. 
For this purpose, two settings are considered for the extended ML model: one where summary statistics of all 55 meteorological variables are candidate predictors, and one where summary statistics of the 55 variables and the 80 spatially aggregated variables are available as candidate predictors. See \cite{github}, \cite{SchulzLerch2022} and \cite{PrimoEtAl2024} for more details on the variables.

\subsection{Post-processing methods and setup}

In the example we present here, the ensemble model output statistics (EMOS, \citealp{gneiting2005calibrated}) post-processing method is compared with its gradient-boosted extension \citep{Messner&2017}. Basic EMOS fits a single parametric predictive distribution using summary statistics from the ensemble forecasts $x_1,\ldots,x_m$, of the variable to be post-processed, where $m=20$ in the CIENS dataset (see Section \ref{sec:NWP}). 
The EMOS model was originally proposed for post-processing forecasts of surface temperature or pressure based on the assumption of a Gaussian distribution. Later, modifications for other weather variables based on suitable distribution assumptions have been proposed (see, e.g., \citealp{ThorarinsdottirGneiting2010, Scheuerer2014, Hemri&2016}).

The EMOS approach can be considered as a special case of a more general distributional regression framework, where a parametric conditional distribution 
\begin{align*}
\mathcal{D}_\theta(Y\,\vert\, X_1=x_1, \ldots, X_m=x_m),
\end{align*}
is assumed for a weather quantity $Y$ given the (potentially vector-valued) ensemble forecasts $X_1=x_1, \ldots, X_m=x_m$.
The distribution parameters $\theta$ are then connected to summary statistics of the ensemble forecasts via pre-specified link functions.
For post-processing wind gusts with EMOS as considered in this example, we assume a logistic distribution truncated in zero
\begin{align*}
\mathcal{D}_{\theta=(\mu,\sigma)}(Y\,\vert\, X_1=x_1, \ldots, X_m=x_m)&=\mathcal{L}_{[0,\infty)}\left(\mu, \sigma \right),
\end{align*}
where $\mu \in\mathbb{R}$ is the location parameter, and $\sigma>0$ the scale parameter of the non-truncated logistic distribution $\mathcal{L}$. The distribution parameters are linked to ensemble forecasts via 
\begin{align*}
\mu&:=a_0+a_1\overline{x},\\
\log(\sigma)&:=b_0+b_1 \log(s),
\end{align*}
where $\overline{x}$ is the empirical ensemble mean, $s$ the empirical ensemble standard deviation, and $a_0,a_1,b_0,b_1\in \mathbb{R}$ are the respective regression coefficients. 
The coefficients are usually estimated by minimizing a proper scoring rule, such as the continuous ranked probability score (CRPS, see Section \ref{sec:verification}).

A key shortcoming of the classical EMOS model is that there is no straightforward way to integrate arbitrarily many input predictors due to a risk of overfitting. In order to allow additional predictors apart from ensemble statistics of the target variable, a gradient-boosted version of EMOS (EMOS-GB) was proposed \citep{Messner&2017}. The model can perform data-driven variable selection by the built-in boosting algorithm and has shown to be largely competitive with other machine learning-based approaches such as random forecasts, see, e.g., \citet{SchulzLerch2022}.

The EMOS-GB setting is analogous to classical EMOS, however, instead of only allowing (summary statistics of) the $m$ ensemble forecasts of the target variable as predictors, a general set of $p$ predictors is considered. These can be (summary statistics of) ensemble forecasts of the variable to be post-processed, and/or (summary statistics of) ensemble forecasts of other weather variables, or additional variables such as spatial and temporal inputs. 
Thus, it is assumed that
\begin{align*}
Y \vert x_1,\ldots, x_p \sim \mathcal{L}_{[0,\infty)}(\mu,\sigma),
\end{align*}
where the location and scale parameter can be linked to all or any arbitrary subset of the available set of $p$ predictors by 
\begin{align*}
\mu &:=a_0+a_1 x_1+\ldots+a_p x_p,\\
\log(\sigma) &:=b_0+b_1 x_1+\ldots+b_p x_p,
\end{align*}
with coefficients $a_i,b_i\in \mathbb{R}$, $i=0,1,\ldots,p$.
The coefficients are initialized as zero and are estimated in an iterative fashion by updating only the single coefficient that improves the current model fit most until a stopping criterion is fulfilled.
This approach leads to an automatic variable selection, as the coefficients of some predictors might never be updated and remain zero, thus playing no role in the resulting model. 

Our implementation of all EMOS and EMOS-GB model variants is based on the \texttt{R} package \texttt{crch} \citep{messner2016heteroscedastic}. This package includes implementations of EMOS and EMOS-GB for several distributions, along with their truncated and censored versions. 
The target variable to be post-processed is wind gust, which is non-negative. In line with \cite{SchulzLerch2022}, the distribution choice for all considered EMOS models is therefore the zero-truncated logistic distribution. 
For the tuning parameters in the boosting algorithm of EMOS-GB we directly follow the implementation in \cite{SchulzLerch2022} and do not conduct an own hyperparameter tuning, i.e., the step length (i.e. the size of the step in the direction of the steepest descent of the gradient of the loss function)  for the coefficient update in the optimization process was set to $\nu=0.05$, the number of boosting iterations to $\text{maxit}=1\,000$, and the stopping criterion was chosen to be the Akaike information criterion (AIC). 

For the basic EMOS model, only the mean of the wind gust ensemble forecasts is used as predictor for the location parameter, and only the standard deviation of the wind gust ensemble forecasts as predictor for the scale parameter. 
As noted above, two versions of EMOS-GB are considered. In the first version mean and standard deviation of the ensemble forecasts of all 55 meteorological variables (including wind gusts) are considered as candidate predictors in the boosting procedure for both the location and scale parameter. This model version will be referred to as EMOS-GB. The extended version, which we will refer to as EMOS-GB-SP, additionally incorporates mean and standard deviation of all 80 spatially aggregated variables as predictors for both parameters. 
The use of these additional input predictors comes with a notable increase in data size (see Table \ref{tab:datasets}).

Separate models are fitted at each of the 170 stations individually using only forecast and observation data from that station. 
Data from the years 2010 - 2020 is used as training period for model fitting. The remaining data from the years 2021 - 2023 is utilized as independent testing period.

\subsection{Forecast verification} \label{sec:verification}

In the presented post-processing example, probabilistic forecast verification is performed via proper scoring rules $S(F,y)$. Such a scoring rule $S$ assigns a value in $\mathbb{R}$ to the pair $(F,y)$ \citep{gneiting2007strictly}, where $F$ is the predictive probability distribution in form of the cumulative distribution function (CDF), and $y \in \mathbb{R}$ a verifying observation. 
A popular scoring rule is the  \textit{continuous ranked probability score} (CRPS; \citealp{Matheson1976}) defined as
\begin{align}
\text{CRPS}(F,y) :=& \int\limits_{-\infty}^{\infty}(F(z)-\mathbb{1}\{z\geq y\})^2\, \mathrm{d} z,\label{eq:CRPS1}
\end{align}
where $F$ is the predictive CDF, $y \in \mathbb{R}$ is the validating observation and $\mathbb{1}$ denotes the indicator function. 
The CRPS can alternatively be represented  in the form 
\begin{align}
\textup{CRPS}(F,y) =& E_F | X - y | - \frac{1}{2} \; E_F | X - X' |,
\end{align}
where $X, \, X' \sim F$ are independent random variables with distribution $F$. 

The relative improvement of a probabilistic forecast $F$ against a reference forecast $F_\text{ref}$ can be quantified via a skill score. 
For the CRPS, the \textit{continuous ranked probability skill score} (CRPSS) is defined as 
\begin{align*}
\text{CRPSS}:=1-\frac{\overline{\text{CRPS}}_F}{\overline{\text{CRPS}}_{F_\text{ref}}},
\end{align*}
where $\overline{\text{CRPS}}_F$ is the mean CRPS of the forecast $F$ (in our example computed over the time range of a station) and $\overline{\text{CRPS}}_{F_\text{ref}}$ denotes the mean CRPS of the  reference forecast $F_\text{ref}$.

All score computations for model fitting as well as validation purposes in the presented example are based on the \texttt{R} package \texttt{scoringRules} \citep{JordanEtAl2019}. 
Apart from the CRPS that evaluates the probabilistic performance, the mean absolute error (MAE) and mean squared error (MSE) were used to assess the deterministic forecast obtained from the probabilistic one by using the median or the mean of the predictive distribution, respectively. Furthermore, the width and the coverage of central prediction interval of the predictive distributions were considered. 
We here follow \citet{SchulzLerch2022} and compute width and coverage of the central prediction intervals, which correspond to a nominal coverage of $(m-1)/(m+1)$ in the case of an $m$-member ensemble, based on the corresponding quantiles of the continuous forecast distributions, or the smallest and largest ensemble member prediction, respectively.

\subsection{Illustrative results}

Table \ref{tab_scores} summarizes the overall predictive performance of the raw ensemble and the considered post-processing methods. One station was removed for the analysis due to extreme outliers in single cases, which affected the performance of some of the investigated methods. 
Clear differences between the forecasting methods can be observed in terms of all considered verification metrics. All post-processing approaches considerably improve the raw ensemble forecasts, and in terms of the CRPS, relative improvements of around 5\% can be observed when including more input predictors, both when transitioning from EMOS to EMOS-GB, as well as from EMOS-GB to EMOS-GB-SP. 

The corresponding deterministic forecasts show similar relative improvements in terms of the MAE and the MSE. 
While the raw ensemble produces forecasts with the narrowest prediction intervals, they lack calibration. 
All post-processing methods show prediction interval coverage values much closer to the nominal coverage.  
Overall, the results shown in this example are largely in line with those reported in \citet{SchulzLerch2022}, who investigated more lead times and a different set of post-processing methods.

\begin{table}
	\caption{Verification scores (mean CRPS, MAE of the median forecast, and MSE of the mean forecast) aggregated over 169 stations and all time points in the test data. Width and coverage are computed for central prediction intervals with a nominal coverage corresponding to a 20-member ensemble ($\approx$ 90.48\%).}\label{tab_scores}%
	\centering 
	\begin{tabular}{@{}lccccc@{}}
		\toprule
		Method  & $\mathrm{CRPS}$ & $\mathrm{MAE}$ & $\mathrm{MSE}$ & $\mathrm{Width}$ & $\mathrm{Coverage}$ \\
		\midrule
		Raw Ensemble & 1.120 & 1.465 & 3.786 & 3.910 & 0.679 \\ 
		EMOS   & 0.972 & 1.359 & 3.447 & 5.963 & 0.920 \\
		EMOS-GB   & 0.928 & 1.298 & 3.114 & 5.301 & 0.894 \\
		EMOS-GB-SP & 0.878 & 1.227 & 2.795 & 5.046 & 0.895 \\
		\bottomrule
	\end{tabular}
\end{table}

\begin{figure}
	\centering
	\includegraphics[width=\textwidth]{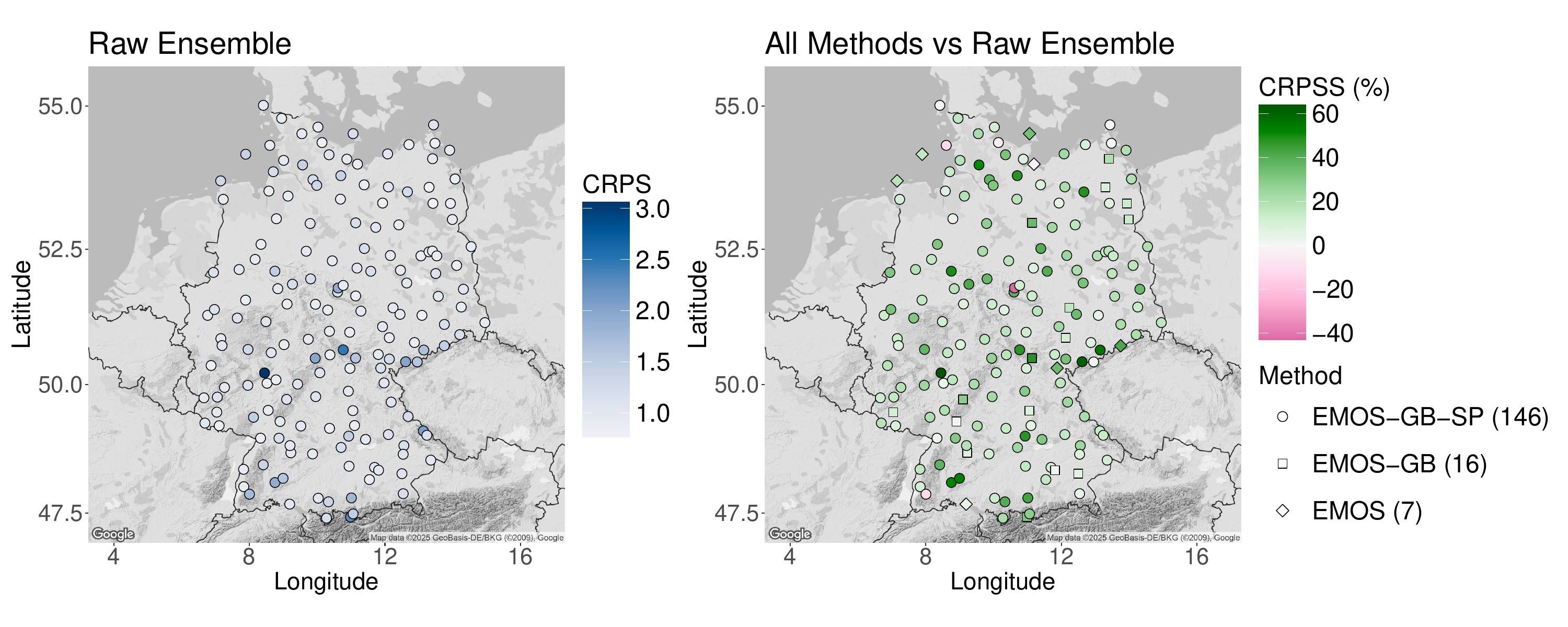}
	\caption{Mean CRPS of the raw ensemble (left) and CRPSS of the best-performing post-processing method using the raw ensemble as a reference model (right) at each station. The different symbols indicate which method performs best at the specific station.}
	\label{fig_meancrps_crpsscomparsions}  
\end{figure}

To illustrate the spatial behavior of the CIENS dataset, Figure \ref{fig_meancrps_crpsscomparsions} provides a more nuanced picture of location-specific differences of the predictive performance. 
The CRPS of the raw ensemble shows values of around $1 \frac{m}{s}$ at most observation stations, but there are a few stations with notably higher mean CRPS values. In particular, those include stations at higher altitudes (see Figure \ref{fig_stations}), where the station altitude might be poorly represented by the model grid, or where the wind velocities are higher and with them their errors.

\begin{figure}
	\centering
	\includegraphics[width=\textwidth]{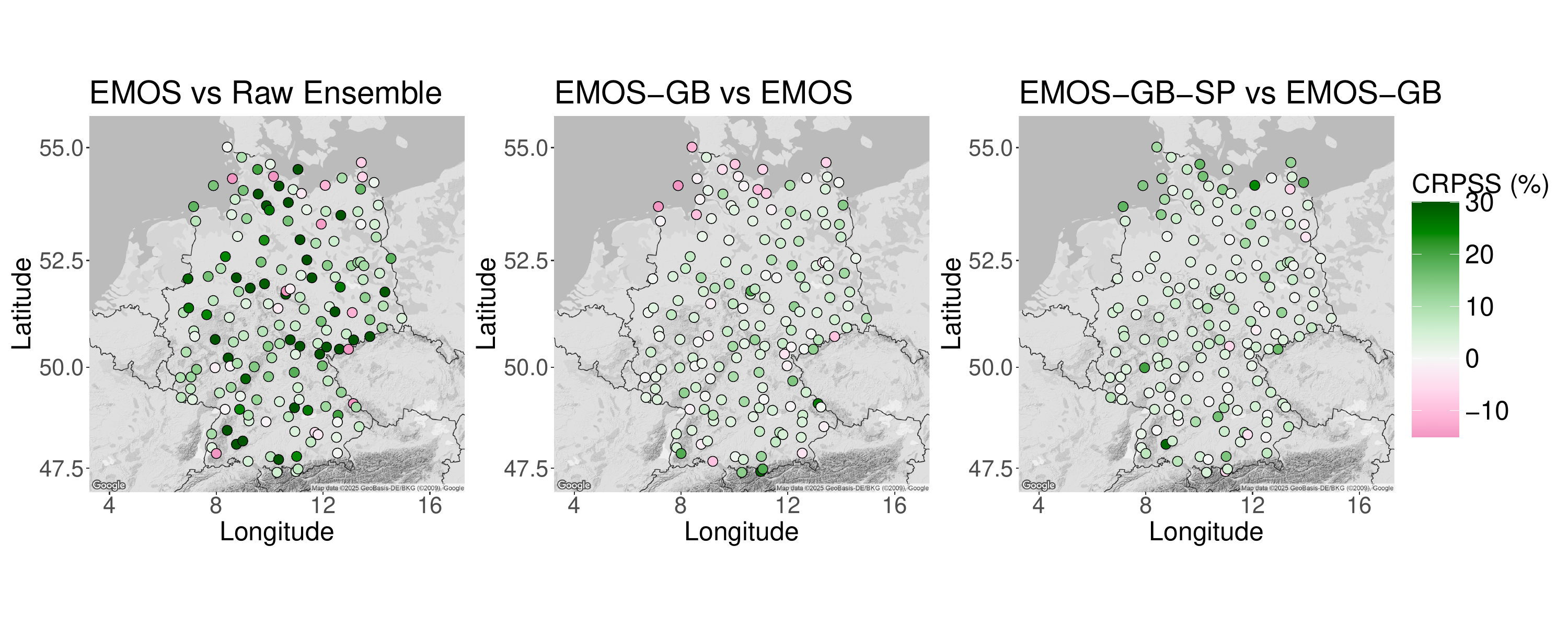}
	\caption{Relative improvements between selected pairs of forecasting methods in terms of the mean CRPS at the individual station locations. The methods mentioned last in each plot title are used as reference for the corresponding CPRSS calculations, respectively. To improve visual clarity, the color scales are censored at $-15\%$ and 30\%, and smaller (resp.\ larger) values are set to $-15\%$ (resp.\ 30\%).}
	\label{fig_crpss_3models}  
\end{figure}

While post-processing improves the raw forecasts at almost all locations, the extent of the improvement of the best-performing method varies strongly. The largest relative improvements of up to around 60\% in terms of the mean CRPS can in particular be observed at some of the locations where the raw ensemble showed the largest mean CRPS values. Furthermore, the EMOS-GB-SP model utilizing not only the meteorological variables but also the spatially interpolated variables as potential input candidates performs best at the highest number of stations. This gives an  indication that including spatial variables or the use of more variables than only the forecast of the variable to be post-processed can provide additional benefit in performance. 
To investigate the effects of including (summary statistics of) NWP forecasts of more meteorological variables in more detail, Figure \ref{fig_crpss_3models} shows relative improvements when using EMOS in comparison to the raw ensemble, and for transitioning from EMOS to EMOS-GB, and from EMOS-GB to EMOS-GB-SP. The latter step corresponds to using the additional spatial inputs which are spatial averages of forecast variables over surrounding areas around the stations. 
The largest relative improvements can be observed for the comparison of EMOS to the raw ensemble forecasts, indicating that already applying a relatively simple post-processing method can yield clear improvements over the raw ensemble.
The use of additional input variables beyond forecasts of the target variable alone leads to further improvements of the predictive performance, as indicated by the largely positive CRPSS values of EMOS-GB at most stations when using EMOS as a reference method. In particular, larger improvements can be observed at higher-altitude stations, or in stations with high observed wind velocities, like coastal stations (e.g. in Northern Germany).

Finally, the additional use of spatial inputs via the EMOS-GB-SP method further improves the predictive performance and shows positive CRPSS values when compared against EMOS-GB at almost all stations. Noticeably, particularly strong improvements can be found for several of the stations at the North Sea coast, for which EMOS-GB without the spatial inputs was not able to improve the performance over EMOS.

\begin{figure}[p]
	\centering
	\includegraphics[width=\textwidth]{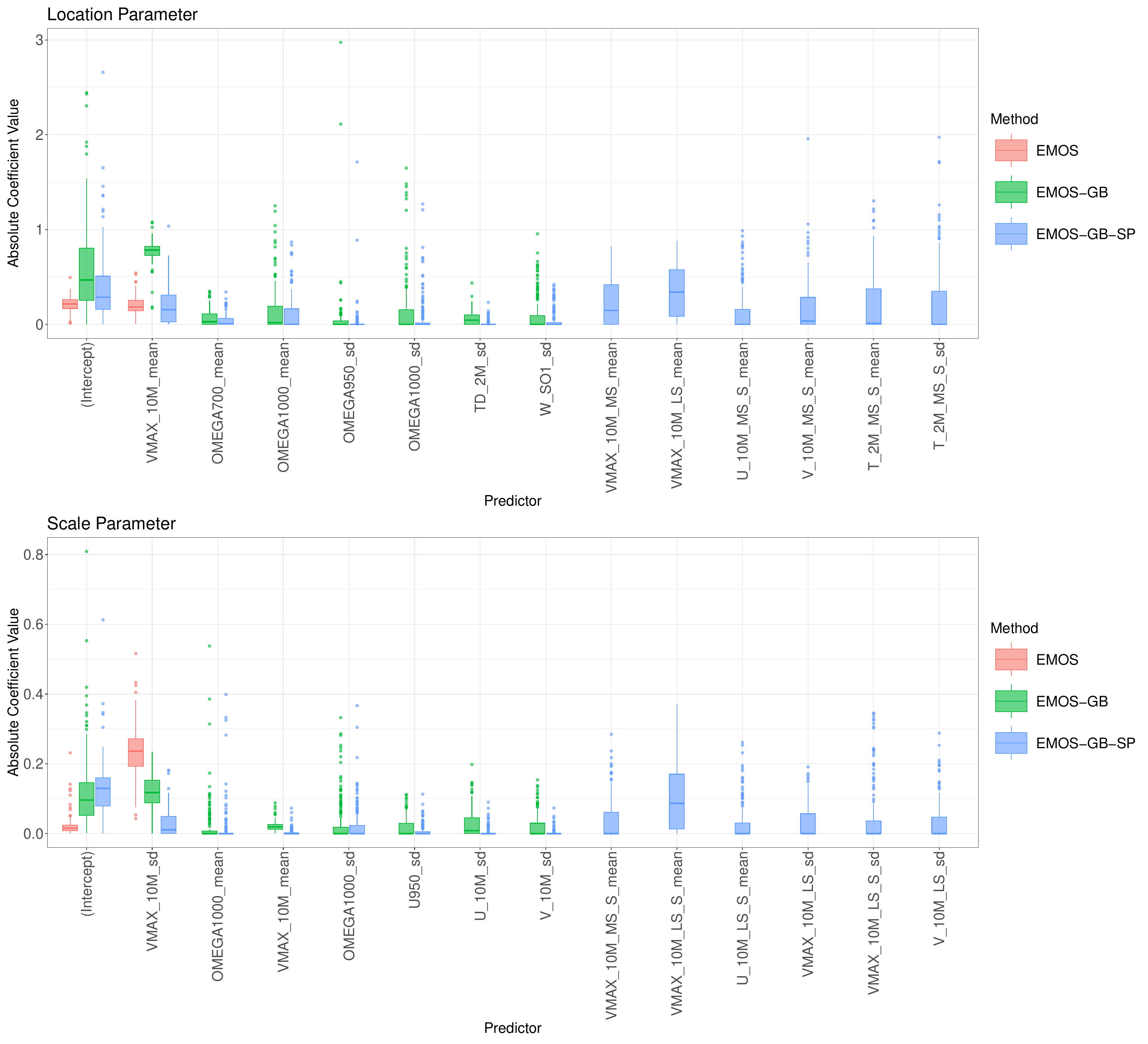}
	\caption{Boxplots summarizing the absolute values of the estimated coefficients of selected input predictors for the location (top) and scale (bottom) parameter of the three post-processing models at the different observation stations. The variables shown here correspond to wind gust forecasts and the most important input predictors for EMOS-GB, and the seven most important input predictors for EMOS-GB-SP (without the previously selected ones). The selection of the most important predictors is based on the mean absolute coefficient values over the stations. The coefficient values for the intercept shown here are divided by five to improve visual clarity.}
	\label{fig_importance}  
\end{figure}

Given the sequence of improvements when including more input predictors, a natural question is which inputs are the most important ones to the different post-processing models. To investigate the predictor importance, Figure \ref{fig_importance} shows boxplots summarizing the distributions of the estimated coefficient values for the location and scale parameters at the different stations for the three post-processing models and selected input variables. Since all inputs are standardized, larger absolute coefficient values indicate a more pronounced influence of that input on the corresponding distribution parameter. By construction, the EMOS and EMOS-GB models only utilize a subset of the available inputs, the number of boxes shown in the figure thus differs for the different methods. 
While there is a large variability across the locations, as indicated by the relatively wide boxes, the intercept and the wind gust predictors consistently are selected as the most important predictors across the three methods. The EMOS-GB model additionally seems to mostly focus on inputs related to the mean and standard deviation of ensemble forecasts of the vertical velocity (or constant) at different pressure levels for the location parameter, and the wind speed in $U$ and $V$ direction for the scale parameter. With the estimated coefficient values for most of those inputs being notably smaller, the EMOS-GB-SP model particularly considers the spatially averaged mean and standard deviation of the wind gust forecasts, along with forecasts of temperature (for the location parameter) and the wind speed in $U$ and $V$ direction (for the scale parameter).

\section{Discussion and conclusions} \label{sec:conclusions}  

We introduce the CIENS dataset, which encompasses more than 12 years of ensemble predictions from DWD's operational weather prediction model, paired with observations of six meteorological variables at 170 weather stations. 
The substantial data volume, particularly the wide range of meteorological variables available in the ensemble predictions, makes it a valuable resource for benchmarking existing methods and developing new statistical and ML methods for ensemble post-processing.
The dataset is structured to facilitate addressing diverse research questions and allows users to extract relevant subsets with minimal effort. 
Accompanying code with example pre-processing functionalities and implementations of selected post-processing methods aims at promoting reproducibility and streamline the dataset's future use.

While a single benchmark dataset cannot capture all aspects relevant to the development of post-processing models, the CIENS dataset offers valuable resources within certain constraints. 
For example, the operational convection-permitting ensemble prediction system at DWD is limited to forecast lead times of up to 21 hours, which may not meet the requirements of all applications.
Further, the included observation stations were selected to focus on user-relevant variables and to ensure consistent coverage over the dataset's time span with minimal data gaps. This necessitated the exclusion of certain variables, such as solar irradiance and visibility, which have been investigated in recent post-processing research \citep{Schulz2021, baran2023statistical, Horat24}.
Another active area of research involves spatial post-processing methods that utilize two-dimensional gridded forecasts as inputs, often leveraging convolutional neural networks \citep{gronquist2021, Veldkamp2021wind, Chapman2022, Li2022precip, horat_lerch_2024_deep}. The CIENS dataset does not include gridded ensemble predictions; instead, in addition to the nearest grid point predictors, it also provides spatial predictors as averages and standard deviations computed over a small set of surrounding grid points for each ensemble member. 
Consequently, the dataset is less suited for developing spatial post-processing models compared to other available benchmark datasets such as EUPPBench \citep{DemaeyerEtAl2023} or WeatherBench 2 \citep{WB2}. Nevertheless, these two datasets do not include any convection-permitting forecasts so far. Therefore, activities are ongoing to extend EUPPBench by a gridded dataset of COSMO forecasts.

Ultimately, the scientific value of a benchmark dataset is determined by its adoption and use. We believe the CIENS dataset has significant potential as a resource for research projects and teaching across disciplines. Its name reflects this ambition, with the acronym derived from the Latin term \textit{ciens}, which loosely translates to ``to put in motion''.

\section{Code and data availability}\label{sec:code_data}

The CIENS data is available from the KITOpen repository at \url{https://dx.doi.org/10.35097/EOvvQEsgILoXpYTK} under a CC BY 4.0 license \citep{CIENS_data_ref}. Exemplary code for the \texttt{R} programming language and additional documentation, along with all code to reproduce the results and usage examples from Section \ref{sec:examples} is available at \url{https://github.com/slerch/CIENS/}.

\section*{Acknowledgments}

Sebastian Lerch and Benedikt Schulz gratefully acknowledge funding from the  German Research Foundation (DFG) within the project C5 ``Dynamical feature-based ensemble postprocessing of wind gusts within European winter storms'' of the Transregional Collaborative Research Center SFB/TRR 165 ``Waves to Weather''. 
Further, support by DFG Grant Number 395388010 (Sebastian Lerch and Annette Möller) and Grant Number 520017589 (Annette Möller) is gratefully acknowledged.
Sebastian Lerch gratefully acknowledges support by the Vector Stiftung through the Young Investigator Group ``Artificial Intelligence for Probabilistic Weather Forecasting''.
We further acknowledge support by the German Weather Service (Deutscher Wetterdienst, DWD) through the SPARC-ML project within the extramural research program, funding reference number 4823EMF01.
We thank Robert Redl for assistance in data handling and the KITOpen team for technical support in making the data publicly available.

\bibliographystyle{myims2}
\bibliography{bibliography.bib}

\end{document}